\documentstyle[prl,aps,epsf]{revtex}
\begin{document}
\title{Understanding Search Trees via Statistical Physics}

\author{Satya N. Majumdar $^{1,2}$, David S. Dean $^2$ and P.L. Krapivsky $^3$}
\address{
{\small $^1$Laboratoire de Physique Th\'eorique et Mod\`eles Statistiques,
        Universit\'e Paris-Sud. B\^at. 100. 91405 Orsay Cedex. France}\\
{\small $^2$Laboratoire de Physique Theorique (UMR C5152 du CNRS), Universit\'e Paul
        Sabatier, 31062 Toulouse Cedex. France}\\
{\small $^3$ Center for Polymer Studies and Department of Physics, Boston 
University, Boston, Massachusetts 02215, USA}
}

\date{\today}

\maketitle

\begin{abstract}
We study the random $m$-ary search tree model (where $m$ stands for the number of branches 
of a search tree), an important problem for data storage in
computer science, using a variety of statistical physics techniques that allow us to
obtain exact asymptotic results. In particular, we show that the probability distributions
of extreme observables associated with a random search tree such as the height and the balanced height 
of a tree have a traveling front structure. In addition, the variance of the number of nodes needed
to store a data string of a given size N is shown to undergo a striking phase transition at a critical value of the 
branching ratio $m_c=26$. We identify the mechanism of this phase transition, show that it is generic 
and occurs in various other problems as well. New results are obtained when each element of the data string 
is a D-dimensional vector. We show that this problem also has a phase transition at a critical
dimension, $D_c= \pi/\sin^{-1}\left(1/\sqrt{8}\right)=8.69363\dots$. 

\end{abstract}

\medskip\noindent {Keywords: Search trees, Fragmentation, Traveling Fronts, Phase Transition}

\medskip\noindent {PACS numbers: 02.50.-r, 89.20.Ff, 89.75.Hc}

\section{Introduction}

\begin{figure}[htbp]
\epsfxsize=8cm
\centerline{\epsfbox{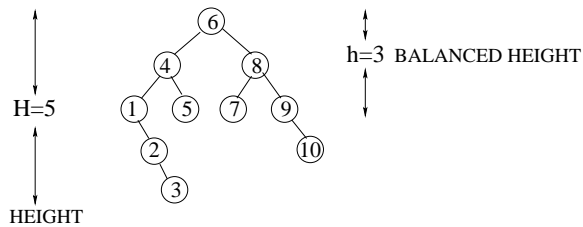}}
\caption{The binary search tree associated with the data string  $\{6,4,5,8,9,1,2,10,3,7\}$. }
\label{fig:bst}
\end{figure}

`Search Trees' are the objects of key interest in an important area of computer science
called `Sorting and Searching' \cite{Knuth} which deals with the basic question: How does one 
store the incoming data to a 
computer in an efficient way so that one spends the minimum time in searching a given data element
if required later? Amongst various search algorithms, the tree based sorting and search algorthims turn out to be 
the most efficient ones. One of the simplest such algorithms is the so called `binary search algorithm' (BSA)
which can be understood by the following simple example. Consider a data string consisting of $N$ elements
which are labelled by the $N$ integers: $\{1,2,\dots,N\}$. These could be the months of the year or the names of people 
etc. Let us assume that this data appears in a particular order, say 
$\{6,4,5,8,9,1,2,10,3,7\}$ for $N=10$ integers. This data is first stored on a binary tree following the simple 
dynamical rule: the first element $6$ is stored at the root of the tree (see Fig. \ref{fig:bst}). The next element in 
the string is 
$4$. We compare it with $6$ at the root and since $4<6$, we store $4$ in the left daughter node of the root.
Had it been bigger than the root $6$, we would have stored it in the right daughter node. 
The next element in the string is $5$. We again start from the root, see that $5<6$, so we 
go to the left branch. There we encounter $4$ and we find $5>4$, so we go the right daughter node of $4$.
This process is continued till all the $N=10$ elements are assigned their nodes and we get a unique binary
search tree (BST) (see Fig. \ref{fig:bst}) for this particular data string $\{6,4,5,8,9,1,2,10,3,7\}$.

Once the data is stored on the tree, it takes very little time to search a required element.
For example, suppose we are looking for the element $7$. We start from the root and comparing 
with $6$ at the root, we know that $7$ must be on the right branch since $7>6$. We then go down one level
and next compare $7$ with $8$ (see Fig. \ref{fig:bst}) and since $7<8$, we look in the left subtree below $8$ and 
immediately find 
$7$. Thus, by construction, we eliminate searching one half of the subtrees at every level. This makes
the search process very efficient. In fact, typical search time to find an element
is $t_{\rm search}= D$ where $D$ is the depth of the element in the tree. Since, roughly speaking, $2^D\sim N$,
one gets $t_{\rm search} \sim O(\log N)$, which is far better than linear search that takes $t_{\rm search}\sim O(N)$.

\begin{figure}[htbp]
\epsfxsize=8cm
\centerline{\epsfbox{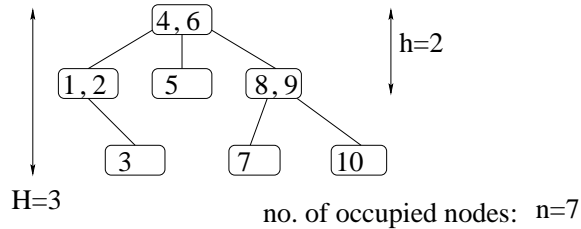}}
\caption{The $m=3$ search tree associated with the data string  $\{6,4,5,8,9,1,2,10,3,7\}$. }
\label{fig:m1}
\end{figure}

An immediate generalization of a BST is an $m$-ary search tree where the tree has $m$ branches. The BST
corresponds to $m=2$. An $m$-ary search tree is constructed in the following way. Each node of the tree
can now hold at most $(m-1)$ elements. One first 
collects the
first $(m-1)$ elements of the data string and stores them together in the root of the tree in an ordered
sequence $x_1<x_2\dots < x_{m-1}$ [see Fig. \ref{fig:m1} for $m=3$]. Next when the $m$-th element arrives, one compares
it first with $x_1$. If $x_m<x_1$, the new element $x_m$ is assigned to the leftmost daughter node of the root.
If $x_1<x_m<x_2$, $x_m$ goes to the daughter node in the second branch and so on. Each subsequent incoming element is 
assigned to either of the $m$ branches according to this above rule. As an example, the same data string
$\{6,4,5,8,9,1,2,10,3,7\}$ of size $N=10$ is stored on a $m=3$ tree in Fig. \ref{fig:m1}. Note that for $m>2$, some of
the nodes of the tree are saturated to their capacity, i.e., are fully occupied with $(m-1)$ elements, while some others 
are only partially occupied.

Once an $m$-ary tree is constructed, one can define a number of observables associated with the tree
which provides information about the structure of the tree. The knowledge of how these observables depend on the data 
size $N$ is of central importance in `sorting and searching'. Amongst many observables, we focus here on $3$ central
objects: 
\begin{enumerate}
\item the height $H_N$ of the tree which is defined as the distance of the farthest node from the root.
For example, in Fig. \ref{fig:m1}, we have $H=3$. The height $H_N$ measures the maximum possible time to search an
element, i.e., it is a measure of the {\it{worst case}} scenario.

\item the balanced height $h_N$ of the tree, defined as the maximum depth upto which the tree is
balanced, i.e., all the nodes upto that level are at least partially occupied. In the example of 
Fig. \ref{fig:m1} we have $h=2$.  Balancing a tree is important for optimizing search algorithms
and hence $h_N$ is an important observable.

\item the number of non-empty nodes $n_N$ of the tree which tells us how many nodes typically one needs
to store a data of size $N$. For example, in Fig. \ref{fig:m1},   
one has $n=7$. Note that for the binary case $m=2$, one has trivially $n_N=N$ since each node can contain
only one element. However, for $m>2$, $n_N$ becomes a variable since some of the nodes may only
be partially filled.

\end{enumerate}

\begin{figure}[htbp]
\epsfxsize=8cm
\centerline{\epsfbox{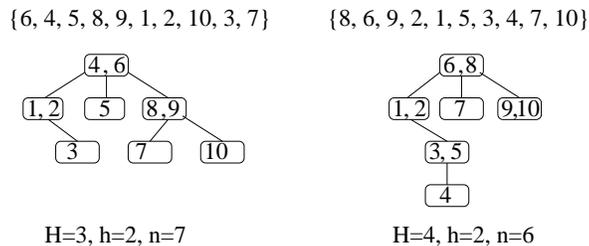}}
\caption{The $m=3$ search trees associated respectively with the data strings  $\{6,4,5,8,9,1,2,10,3,7\}$
and \{8,6,9,2,1,5,3,4,7,10\}. }
\label{fig:m2}
\end{figure}

Usually the data arrives at a computer in random order. To study this situation, one considers the simplest model 
called the `random $m$-ary search tree model' (RmST) where one assumes that the incoming data string can arrive in
any of the $N!$ possible order or sequence, each with equal probability. For each of these sequences, one has
an $m$-ary tree and the associated observables $H_N$, $h_N$ and $n_N$. As the sequence changes, the corresponding
tree changes and hence these observables also take on different values. 
For example, in Fig. \ref{fig:m2}, we show two sequences, their corresponding $m=3$ trees and the values
of the $3$ observables.
The central question of importance is:
given that all the $N!$ sequences occur with equal probability, what are the statistics of $H_N$, $h_N$ and
$n_N$? For example, what are the averages, variances or even the full probability distributions of these observables?   

The statistics of $H_N$ and $h_N$ have been studied by computer scientists over the past two decades and many 
nontrivial results have been found\cite{Robson,FO,Dev1,Mahmoud}. For example, the
average height and the average balanced height of a random $m$-ary search tree have the following asymptotic behaviors 
for large $N$,
\begin{eqnarray}
\langle H_N\rangle &\approx& a_m \log N + b_m \log(\log N) +\dots \nonumber \\
\langle h_N\rangle &\approx& c_m \log N + d_m \log(\log N) +\dots.
\label{asymp1}
\end{eqnarray}  
While the leading $\log (N)$ behavior was proved by Devroye\cite{Dev1} who also computed
the coefficients $a_m$ and $c_m$, the subleading double logarithmic behavior was conjectured only recently 
by Hattori and Ochiai\cite{HO}, who found $b_2\approx -1.9 $ numerically. Also, the variance and even the higher moments 
of $H_N$ and $h_N$ were found to be independent of $N$ for large $N$\cite{Robson2,Drmota}.
 
The study of the statistics of $n_N$, on the other hand, is relatively recent\cite{MP,CH,CP}. Chern and Hwang recently 
found\cite{CH}
that while the average $\mu_N = \langle n_N\rangle \sim N$ for large $N$ (as one should expect), the variance
$\nu(N)= \langle \left(n_N- \mu(N)\right)^2\rangle$ undergoes a striking {\it{phase transition}} as a function of $m$.
They found that
\begin{eqnarray}
\nu(N) &\sim & N \quad\quad {\rm for}\quad  m\le 26 \nonumber \\
       &\sim & N^{2\theta(m)} \quad\,\, {\rm for} \quad  m> 26,
\label{asymp2}
\end{eqnarray}
where the exponent $\theta(m)>1/2$ depends on $m$ for $m>26$. 

The various important results mentioned above were derived by the computer scientists using sophisticated
probabilistic methods which, though rigorous, are often not simple. As physicists, one would like
to understand and derive these results in a physically transparent way. Moreover, as it often happens,
a physical approach has the advantage that it can make links with other problems and also the 
generalization
often becomes easier. In a series of recent papers\cite{KM,MK1,BKM,DM}, we were able to build up a statisical
physical approach to the RmST problem which not only allowed us to rederive many asymptotically exact results 
(known previously only via
rigorous probabilitic methods) in a physicaly transparent way, but also led to many new results,
generalizations and links to other problems. For example, we were able to generalize our results to
other search trees such as the `digital search trees' (DST) (which has links to the 
Lempel-Ziv data compression algorithm) and found an exact mapping 
between the DST and the problem of the directed diffusion limited aggregation (DLA) problem on the Bethe 
lattice\cite{SM}.
The latter problem was first studied by Bradley and Strenski numerically\cite{BS} and 
remained unsolved for many years.
Our approach provides an exact asymptotic result for this DLA problem\cite{SM}.

Our strategy was to first map the RmST problem to a
random fragmentation problem which was more amenable to statistical physical analysis. The main new discovery
was that the distributions of the height $H_N$ and the balanced height $h_N$, which are `extreme' variables, 
have a `traveling front' structure. The `traveling fronts' appear in many physics and biology problems
and have been well studied over the past few decades\cite{TF}. The techniques developed
in analysing traveling fronts were then useful to derive many asymptotically exact results for the RmST
problem. Subsequently, we found that in many problems where one is interested in finding the statistics of extreme 
variables, there is often a `traveling front' structure\cite{TF1,TF2}. 

For the number of non-empty nodes $n_N$, which is 
not an extreme variable, a different statistical physics approach (equivalent to a backward Fokker-Planck method)
was used which allowed us to understand the mechanism of the phase transition, the significance of the
critical number $26$ and calculate the exponent $\theta(m)$ exactly\cite{DM}. We were also able to
show that this phase transition is rather generic and occurs in other problems as well. Our approach allowed us 
to generalize to the case when the data string consists of $N$ $D$-dimensional vectors. For example, we found that
there is again a phase transition at a critical dimension $D_c= \pi/\sin^{-1}\left(1/\sqrt{8}\right)=8.69363\dots$.
In the next few sections we outline our approach and state the main results.  

\section{Mapping to a Fragmentation Process}

\begin{figure}[htbp]
\epsfxsize=8cm
\centerline{\epsfbox{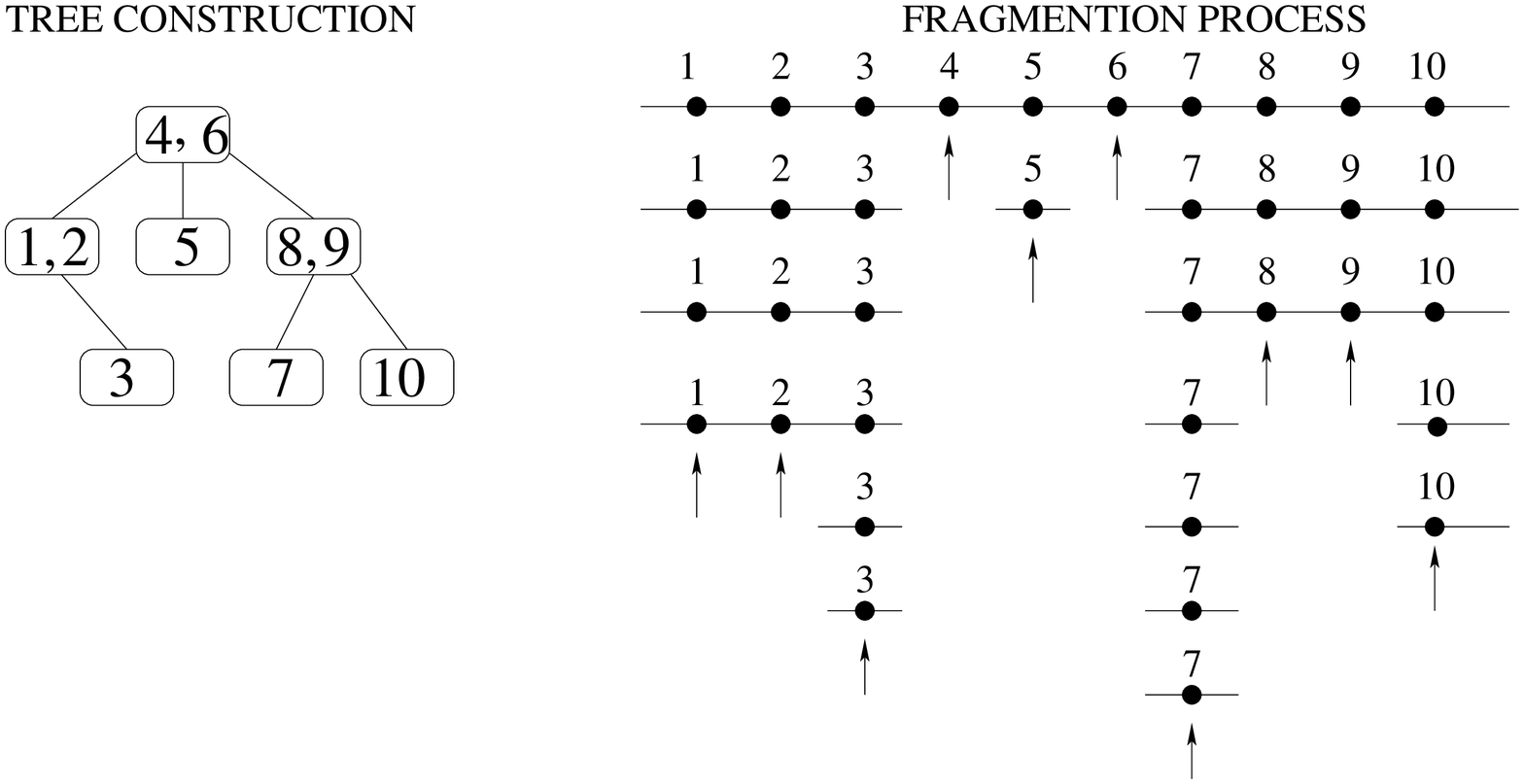}}
\caption{The $m=3$ search tree associated with the data string  $\{6,4,5,8,9,1,2,10,3,7\}$
and the corresponding fragmentation process.}
\label{fig:m3}
\end{figure}

Our strategy is to first map the problem of RmST to a random fragmentation problem\cite{MK1,DM}, which 
in some sense, is more familiar to physicists. This fragmentation procedure can then be viewed 
as a dynamical process and one can write down its evolution equation fairly easily. 
This mapping is best understood in terms of an example.
Let us take our favorite data string $\{6,4,5,8,9,1,2,10,3,7\}$ and store it on an $m=3$ tree as in
Fig. \ref{fig:m1} and also shown in the left half of Fig. \ref{fig:m3}.  
In the fragmentation problem, one starts with a stick (or interval) of length $N=10$. Once the first two elements
$4$ and $6$ are stored in the root of the tree, the remaining elements will belong either to the interval
$[1-3]$, $[5]$, or $[7-10]$, which are subsequently completely disconnected from each other.  Thus storing the first 
two elements is equivalent, in the fragmentation problem, to breaking the original interval $[1-N]$ of length $N$ into
$3$ smaller intervals $[1-3]$, $[5]$, and $[7-10]$. The two break points $4$ and $6$ are chosen uniformly from the
$N$ points $\{1,2,3\dots,N\}$ in the RmST problem (shown by the arrows in Fig. \ref{fig:m3}). Next, when the element $5$ 
arrives in 
the tree, it corresponds to breaking the interval containing the element $5$ randomly into $3$ parts (this breaking
is not 
shown explicitly in Fig. \ref{fig:m3}).
The process is then repeated for other elements. Note that in the fragmentation problem, an interval
breaks {\it{iff}} there is a an element (shown by black dots) inside the interval. Thus there is a threshold phenomenon: 
if the length of a stick is too small so that it doesn't have an element (black dot) in it, one doesn't fragment it any 
more. We denote 
this theshold length by $N_0$ (In our example, $N_0=1$). It just sets the unit of length and its actual
value is not important for the asymptotic large $N$ analysis. 
Those intervals which still have black dots in them (and thus have lengths $> N_0$) are thus `alive' and 
will fragment 
subsequently, but those whose lengths are $<N_0$ are `dead'. Thus, when all the $N$ elements 
are stored in the tree,
all the intervals in the corresponding fragmentation problem become `dead'. 

Note that, in the fragmentation problem, at each step (shown by different levels on the right in Fig. \ref{fig:m3})
there is only one `splitting event'. Each time an interval splits, it corresponds to 
storing in a node on the tree.
Thus completing a tree is equivalent to ending one `history' of the fragmentation process (at the end of which 
all intervals are `dead'). Evidently,
the number of non-empty nodes $n_N$ in the tree is exactly same as the total number of `splitting events'
in the history of the fragmentation process (for example, in  Fig. \ref{fig:m3} the number of nodes
on the tree and the number of splitting events are both $7$). Let $l_i$'s 
denote the lengths of intervals 
in the fragmentation problem at a given stage.
One can then set up a dictionary between the two problems\cite{MK1,DM} and it is easy to see that
\begin{enumerate}
\item{Height $H_N$:} Prob[$H_N < n$]= Prob[ $l_1<1$, $l_2<1$, $\ldots$ after $n$ steps of fragmentation.]
\item{Balanced height $h_N$:} Prob[$h_N >n$]=Prob[ $l_1>1$, $l_2>1$, $\ldots$ after $n$ steps of fragmentation.]
\item{Number of non-empty nodes $n_N$:} Prob[$n_N=n$]= Prob[there are a total of $n$ `splitting events' till the
end of the fragmentation process.]
\end{enumerate}

\section{Analysis of the Fragmentation Problem}

Once this dictionary is set up, one can forget about the original tree problem and focus
on the fragmentation problem. For simplicity, we will also assume that the lengths of sticks in the
fragmentation problem are continuous variables. This is because the original discrete problem and the
continuous problem will have the same asymptotic properties for large $N$, but the continuous problem is easier to 
handle. Thus, in the continuous problem, we start with a stick of length $N$ where $N$ is large. We break
it randomly into $m$ fragments of lengths $r_1N$, $r_2N$, $\dots$, $r_mN$ where the fractions $r_i$'s
are random numbers between $[0,1]$ that satisfy the length conservation condition, $\sum_{i=1}^m r_i =1$.
At this point, we will consider a general problem where the fractions $r_i$'s are drawn from a
normalized joint distribution $\eta(r_1,r_2,\dots, r_m)$. The RmST problem would correspond to a specific choice
of this joint distribution. Note that in the RmST problem, all the $N!$ permutations of the original sequence occur
equally likely. This means that the first $(m-1)$ elements are random, each drawn independently and uniformly from 
$[1-N]$. In the fragmentation language, this means that each of the fractions $r_1$, $r_2$, $\dots$, $r_{m-1}$ is chosen 
from a uniform distribution between $0$ and $1$ and
then one sets, $r_m=1-(r_1+r_2+\dots+r_{m-1})$. This leads to the normalized joint distribution
$\eta(r_1,r_2,\dots, r_m)= (m-1)! \delta(\sum_{i}^m r_i-1)$\cite{MK1}. 
One of the advantages of our method is that it allows us to obtain exact results for arbitrary joint distribution of 
the fraction $r_i$'s, not necessarily only for the uniform case. The RmST problem just corresponds to a special case.
\begin{figure}[htbp]
\epsfxsize=8cm
\centerline{\epsfbox{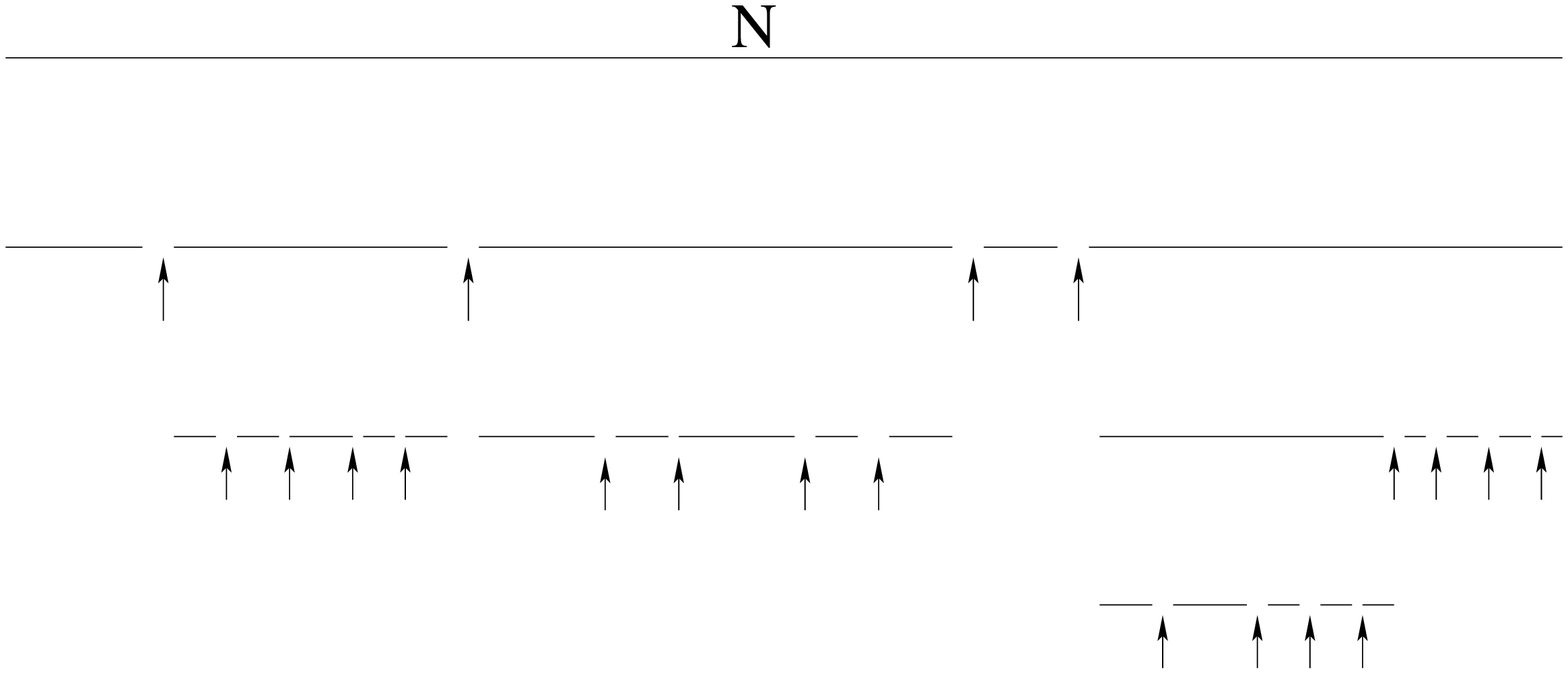}}
\caption{The fragmentation process with continuous lengths for $m=5$. The arrows denote the break points.}
\label{fig:frag1}
\end{figure}

After the first spliting event, we examine the lengths of each of the $m$ fragments. If the length of a
fragment is already less than $N_0=1$, we proclaim it `dead' and it doesn't split any further. Those fragments
with lengths $> N_0=1$ are `alive' and each of those `alive' fragments is further split into $m$ pieces
by drawing, for each piece independently, a set of fractions $r_i$'s from the identical joint distribution
$\eta\left(\{r_i\}\right)=\eta(r_1,r_2,\dots, r_m)$. This process is then repeated till all the 
intervals become `dead', i.e., their lengths
become $< N_0=1$. A pictorial representation is given in Fig. \ref{fig:frag1} with $m=5$.

For subsequent analysis, it is 
useful to define the `marginal'  distribution
$\eta(r_i)$ of any one of the fractions as, 
\begin{equation}
\eta(r_i)=\int \eta\left(\{r_i\}\right)dr_1\dots dr_{i-1} dr_{i+1}\dots dr_m.
\label{ind1}
\end{equation}   
For simplicity, we will assume isotropy, i.e., $\eta(r_i)=\eta(r)$ is independent of the index $i$ and is thus the 
same for each fragment. For example, for the RmST problem, one easily gets\cite{MK1}
\begin{equation}
\eta(r)= (m-1) (1-r)^{m-2}.
\label{ind2}
\end{equation}
for $0\le r\le 1$.
Note that for binary trees $m=2$, where one breaks a stick into two pieces, one gets $\eta(r)=1$ for $0\le r\le 1$, the 
usual uniform distribution for the break point.

\subsection{The Height and the Balanced Height}

\begin{figure}[htbp]
\epsfxsize=8cm
\centerline{\epsfbox{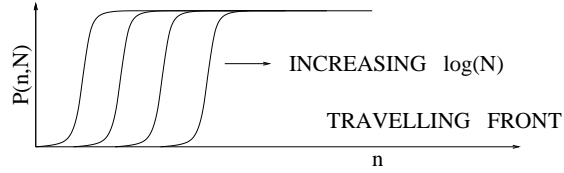}}
\caption{The traveling front structure of the solution of Eq. (\ref{recur1}).}
\label{fig:front1}
\end{figure}

Let us denote the cumulative height distribution ${\rm Prob}[H_N<n]$ by $P(n,N)$. Using 
the dictionary outlined before,
we have $P(n,N)$= ${\rm Prob}$ [$l_1<1$, $l_2<1$, $\ldots$ after $n$ steps of 
fragmentation starting with the initial length $N$] where $l_i$'s are the lengths of the 
intervals. It is then easy to set up a recursion satified by $P(n,N)$ for the 
fragmentation process. Consider the first splitting where we have $m$ new 
intervals
of lengths $r_1N$, $r_2N$, $\dots$, $r_mN$. Each of these new pieces will have 
subsequent histories of evolution completely independent of each other. Hence, it follows
\begin{equation}
P(n,N)= \int \left[\prod_{i=1}^m P(n-1,r_iN)\right] \eta\left(\{r_i\}\right) dr_1 dr_2\dots dr_m,
\label{recur1}
\end{equation}
satisfying the condition, $P(n,1)=1$ for all $n\ge 1$ (this follows from the fact that if the initial 
length is $1$, after the first splitting all the lengths will be $<1$). 
The equation (\ref{recur1}) is reminiscent of a backward Fokker-Planck equation.
It is further useful to
make a change of variables, $t=\log (N)$ and $\epsilon_i=-\log (r_i)$. The joint distribution 
of $\epsilon_i$'s are given by ${\tilde \eta}\left(\{\epsilon_i\}\right)\prod_{i} d\epsilon_i=
\eta\left(\{r_i\}\right)\prod_{i} dr_i$. Then the Eq. (\ref{recur1}) reduces to,
\begin{equation}
P(n,t)= \int \left[\prod_{i=1}^m P(n-1, t-\epsilon_i)\right] {\tilde 
\eta}\left(\{\epsilon_i\}\right)
d\epsilon_1 d\epsilon_2\dots d\epsilon_m.
\label{recur2} 
\end{equation}
The Eq. (\ref{recur1}) (or equivalently Eq. (\ref{recur2}) is nonlinear and hence is difficult to solve 
exactly. However, if one
plots the numerical solution of Eq. (\ref{recur1}), one finds a traveling front structure
as shown in Fig. \ref{fig:front1}. 

This means that the solution at late `times' $t$ has the
structure, $P(n,t)\sim f\left(n-n_f(t)\right)$, where $n_f(t)$ is the position of the front
at `time' $t$. Note that the front retains its shape as $t$ increases which indicates that the
width of the front remains of $O(1)$ even as $t\to \infty$. The front position advances with
a uniform velocity, i.e., $n_f(t)\approx v t$, to leading order for large $t$ where the velocity
$v$ is yet to be determined. We substitute $P(n,t)=1-F(n-vt)$ in Eq. (\ref{recur2}) and then
focus near the large $n$ tail where $F$ is small and hence one can linearize the equation to get
\begin{equation}
F(x)=m\int_0^{\infty} F(x-1+v\epsilon) {\tilde \eta}(\epsilon) d\epsilon,  
\label{recur3}
\end{equation}
where ${\tilde \eta}(\epsilon) d\epsilon=\eta(r)dr$ is the effective induced distribution associated
with any one of the fractions. This linear equation clearly admits an exponential solution
$F(x)= e^{-\lambda x}$ provided $\lambda$ is related to $v$ via the dispersion relation,
\begin{equation}
1= m e^{\lambda} \int_0^{\infty} e^{-\lambda v \epsilon} {\tilde \eta}(\epsilon) d\epsilon.
\label{disp1}
\end{equation}
Thus, in principle, one can have a whole family of possible velocities $v(\lambda)$ parametrized by $\lambda$.
However, in practice, the front has a unique velocity. So, how does one select this unique velocity
from a continuous one parameter family of possible velocities? It turns out that the   
solution $v(\lambda)$ of Eq. (\ref{disp1}) is a nonmonotonic function of $\lambda$ with a single minimum 
at $\lambda=\lambda^*$ that depends on the distribution ${\tilde \eta}(\epsilon)$. According to the velocity selection 
principle developed in the traveling front 
literature\cite{TF,TF2}, the front always chooses this minimum velocity $v(\lambda^*)$ as long as 
the initial condition is sharp enough. Thus the leading front position is given by $n_f(t)\approx v(\lambda^*) t$
where $v(\lambda^*)$ is obtained by minimizing $v(\lambda)$ in Eq. (\ref{disp1}) with respect to $\lambda$. 
Moreover, it turns out that the leading front position has an 
associated slow logarithmic correction\cite{TF},
\begin{equation}
n_f(t)= v(\lambda^*) t - \frac{3}{2\lambda^*} \log (t) +\dots.
\label{fras1}
\end{equation}
Note that since ${\rm Prob}[H_N<n]=P(n,N)$, the expected height $\langle H_N\rangle = \sum_n[1-P(n,N)]
\approx n_f(t)$ where $t=\log (N)$. This follows from the fact that the front rises sharply from $0$
for $n<n_f(t)$ to $1$ for $n>n_f(t)$. Thus, the summation $\sum_n[1-P(n,N)]$ can be replaced by the front 
location $n_f(t)$. 
Using Eq. (\ref{fras1}), we then get
\begin{equation}
\langle H_N\rangle =  v(\lambda^*)\log N - \frac{3}{2\lambda^*} \log \left(\log (N)\right) +\dots
\label{fras2}
\end{equation}
This then provides a physical derivation of the result in Eq. (\ref{asymp1}) where 
we identify the constant $a_m=v(\lambda^*)$
with the velocity of the front and the constant $b_m=-3/{2\lambda^*}$ as the prefactor of the correction term. 
Note that our result is
more general than the RmST (which is just a special case where the break points are chosen uniformly). 
Our derivation also provides a proof for the double logarithmic form of the 
the correction term previously only conjectured by Hattori and Ochiai\cite{HO}. 

For the RmST problem, we have $\eta(r)$ from Eq. (\ref{ind2}). This gives,
${\tilde \eta}(\epsilon)= (m-1)(1-e^{-\epsilon})^{m-2} e^{-\epsilon}$. Substituting this in Eq. (\ref{disp1}), we get the 
dispersion relation,
\begin{equation}
m (m-1) e^{\lambda} B(\lambda v +1, m-1)=1,
\label{disp2}
\end{equation}
where $B(m,n)$ is the standard Beta function. For example, for the binary case $m=2$, one gets from Eq. (\ref{disp2}),
$v(\lambda)= (2e^\lambda -1)/\lambda$ which has a minimum at $\lambda^*= 0.76804\dots$ with $v(\lambda^*)=4.31107\dots$.
One then gets for $m=2$ an exact result,
\begin{equation}
\langle H_N\rangle = 4.31107\dots\log N - 1.95303\dots \log \left(\log (N)\right) +\dots
\label{asymp3}
\end{equation}
Similarly, one can derive the exact asymptotic behavior for all $m$ and for arbitrary fraction distribution 
$\eta(r)$\cite{MK1}. Note that for the binary case $m=2$, the same double logarithmic correction term
was also found by Reed using rigorous probabilistic methods\cite{Reed}, but our results seem to
be more general.

For the balanced height $h_N$, the analysis is similar. The cumulative probability
$Q(n,N)={\rm Prob}[h_N>n]$ satisfies exactly the same recursion relation as in Eq. (\ref{recur1}), except the initial 
condition is different\cite{MK1}. One has, $Q(n,1)=1 $ for $n\le 1$ and $Q(n, 1)=0$ for $n>1$. Again, the solution has a 
traveling front structure, except now it has a $[1-0]$ front as opposed to the $[0-1]$ front in the height case.
Proceeding along the same path, one obtains the asymptotic front position and hence the average balanced height,
\begin{equation}
\langle h_N\rangle =  v(\lambda^*)\log N + \frac{3}{2\lambda^*} \log \left(\log (N)\right) +\dots
\label{bfras2}
\end{equation}
 where $v(\lambda^*)$ is determined by {\it {maximizing}} $v(\lambda)$ obtained from the dispersion relation,
\begin{equation}
1= m e^{-\lambda} \int_0^{\infty} e^{+\lambda v \epsilon} {\tilde \eta}(\epsilon) d\epsilon.
\label{disp3}
\end{equation}
Note that this dispersion relation is the same as in Eq. (\ref{disp1}) provided one changes the sign of $\lambda$.
This reflects the so called `duality' between the height and the balanced height\cite{MK1}. For the $m=2$
binary case, we get from Eq. (\ref{disp3}), $v(\lambda)= (1-2e^{-\lambda})/\lambda$ which has a maximum
at $\lambda^*=1.67835\dots$ and $v(\lambda^*)= 0.373365\dots$. This gives\cite{MK1},
\begin{equation}
\langle h_N\rangle = 0.373365\dots\log N + 0.89374\dots \log \left(\log (N)\right) +\dots
\label{asymp4}
\end{equation}
Note that the sign of the correction term is different in Eqs. (\ref{asymp3}) and (\ref{asymp4}).
Similarly, one can derive exact asymptotic results for all $m$ as well for any arbitrary distribution $\eta(r)$.

\subsection{Number of Non-empty Nodes}

We now turn to the statistics of the number of non-empty nodes $n_N$ required to store a data string of size $N$.
Once again, the fragmentation representation turns out to be useful. One can easily write down a recursion relation for 
$n_N$ by noting that $n_N$ is just the total number of spilitting events in the fragementation process till it stops,
given that it started with an initial stick of length $N$. After the first spliting one has $m$ pieces
of lengths $r_1N$, $r_2N$, $\dots$, $r_mN$ whose subsequent histories are completeley independent of each other.
Note that an interval splits {\it {iff}} its length is $>N_0$ where $N_0$ is the threshold length. Evidently,
if the starting length $N<N_0$, $n_N=0$ since there would not be any splitting. However, if $N>N_0$, one can
write a recursion\cite{DM}, 
\begin{equation}
n_N \equiv n_{r_1N} + n_{r_2N} + \dots + n_{r_m N} +1,
\label{node1}
\end{equation}
where the fractions $r_i$'s are again random numbers satisfying $\sum_{i=1}^m r_i=1$ that are drawn from a
joint distribution $\eta\left(\{r_i\}\right)$. The term $1$ on the right hand side of Eq. (\ref{node1}) just
counts the first splitting and the rest of the terms count the total number of subsequent splitting  
events arising from each of the $m$ pieces generated after the first splitting. The $\equiv$ symbol
represents `equivalence in law', i.e., the left and the right hand side of the $\equiv$ symbol have
the same probability distribution.

Taking average on both sides of Eq. (\ref{node1}), one finds that the average number of nodes or the
`splitting events' $\mu(N)= \langle n_N\rangle$ satisfies an integral equation\cite{DM}
\begin{equation}
\mu(N)= m \int_{N_0/N}^{1} \mu(rN) \eta(r) dr +1.
\label{node2}
\end{equation}
This integral equation can be solved exactly\cite{DM}. One finds that, $\mu(N)= g(N/N_0)$ where the scaling function
$g(z)$ is given by
\begin{equation}
g(z) = \alpha_0 + \alpha_1 z + \sum_{k=2}^{\infty} \alpha_k z^{\lambda_k},
\label{node3}
\end{equation}
where $\lambda_k$'s are the roots of the following equation with $Re(\lambda_k)<1$ ,
\begin{equation}
m \int_0^{1} r^{\lambda} \eta(r) dr=1.
\label{indic1}
\end{equation}
Note that $\lambda=1$ is always a root of Eq. (\ref{indic1}). This follows from
the following observation. By averaging
the sum rule, $\sum_{i=1}^m r_i=1$ one gets $\langle r \rangle=1/m$ which shows
that $\lambda=1$ is always a solution of Eq. (\ref{indic1}). In fact, the linear term
in Eq. (\ref{node3}) corresponds to this root at $\lambda=1$. Furthermore, one can
prove that all the others roots $\lambda_k$'s are complex, if $\lambda_k$ is a root its complex conjugate 
$\lambda_k^*$ is also a root and all these other roots lie in the complex $\lambda$
plane to the left of the imaginary line at $\lambda=1 + iz$, i.e., 
$Re(\lambda_k)<1$. The leading behavior of the average for large $N$ is 
given by the linear term and one
gets, $\mu(N)\sim \alpha_1 N/N_0$ where
\begin{equation}
\alpha_1= -\frac{1}{m \int_0^1 r \log (r) \eta(r) dr}.
\label{node4}
\end{equation}
For the RmST problem, we have $\eta(r)=(m-1)(1-r)^{m-2}$ which gives, $\alpha_1= 1/\sum_{k=2}^m 1/k$.

For the variance $\nu(N)=\langle \left(n_N-\mu(N)\right)^2\rangle$, one can similarly write down a 
recursion relation\cite{DM} starting from Eq. (\ref{node1}),
\begin{equation}
\nu(N)= m \int_{N_0/N}^{1} \nu(rN) \eta(r) dr +J,
\label{node5}
\end{equation}
where $J$ represents a `source' term that depends on the form of the first moment $\mu(N)$. More precisely, if
$S= \sum_{i=1}^m \mu(r_i N)$, then $J= \langle \left( S- \langle S\rangle\right)^2\rangle$. The significant 
fact about this problem is that the equation for the second moment `closes' in the sense that it involves
only second and first moments, but not higher moments. It does not have the usual hierarchy problem
that one often encounters in statistical mechanics problem. This fact makes this problem analytically
tractable. This source term $J$ also turns out to be responsible for driving the `phase transition'
in the variance. This is a new mechanism of phase transition that one has not encountered before in other problems.

Using the exact solution for the first moment $\mu(N)$ from Eq. (\ref{node3}), one can evaluate the source term $J$
which turns out to be only a function of $z=N/N_0$ and for large $z$ one gets, 
\begin{equation}
J(z)\approx \beta_1 z^{2\lambda_2} + \beta_2 z^{2\lambda_2^*} + \beta_3 z^{\lambda_2+\lambda_2^*} + \dots
\label{node6}
\end{equation}
where $\lambda_2$ (and its complex conjugate $\lambda_2^*$) are the nearest zeros of the equation,
$m \int_0^{1} r^{\lambda} \eta(r) dr=1$ to the left of the line $Re(\lambda)=1$ in the complex $\lambda$ plane. 
Substituting this asymptotic behavior of $J(z)$ in Eq. (\ref{node5}) and solving the integral equation, one finds
that $\nu(N)=Y(N/N_0)$ where the asymptotic behavior of $Y(z)$ for large $z$ depends on the value of
$Re(2\lambda_2)$. One finds that as $z\to \infty$, $Y(z)\sim z$ (as in the case of the first moment)
provided $Re(2\lambda_2)<1$. In this case, the source term $J$ turns out to be insignificant and gives rise only 
to subleading correction terms. However, if $Re(2\lambda_2)>1$, the source term $J(z)$ becomes significant and controls 
the 
asymptotic behavior of $Y(z)$ and one gets, $Y(z)\sim z^{2\theta}$ where $\theta= Re(\lambda_2)$. 

Note that the root $\lambda_2$ is a function of $m$. As one tunes $m$, $\lambda_2$ changes but always stays
to the left of the line $Re(\lambda)=1$ in the complex $\lambda$ plane. However, for small $m$, $Re(2\lambda_2)<1$,
i.e., $\lambda_2$ stays to the left of the line $Re(\lambda)=1/2$. Then as $m$ exceeds a critical value $m_c$,
$\lambda_2$ crosses the line $Re(\lambda)=1/2$ from its left to its right and $Re(2\lambda_2)>1$, leading to
a phase transition in the large $N$ behaviour of $\nu(N)$. Thus the critical value of $m_c$ is determined 
from the condition, $Re(\lambda_2)=1/2$. For the RmST problem, substituting $\eta(r)=(m-1)(1-r)^{m-2} $ in Eq. 
(\ref{indic1}) one gets,
$m(m-1)B(m-1,\lambda+1)=0$. One then obtains $\lambda_2$ using the Mathematica. Setting $Re(\lambda_2)=1/2$ determines
the critical value, $m_c= 26.0561\ldots$. Note that, once we have written down the moment equations, $m$ can be 
treated as a continuous parameter, even though in actual search trees $m$ is always an integer. We thus get a 
very general result,
\begin{eqnarray}
\nu(N) &\sim & N \quad\quad {\rm for}\quad  m\le m_c \nonumber \\
       &\sim & N^{2\theta(m)} \quad\,\, {\rm for} \quad  m> m_c,
\label{node7}
\end{eqnarray}
for arbitrary breaking distribution $\eta(r)$ where $m_c$ is determined from $Re(\lambda_2)=1/2$
and $\theta(m)= Re(\lambda_2)$ where $\lambda_2$ is determined from Eq. (\ref{indic1}). 
For the RmST case in particular, we get $m_c=26.0561\ldots$.    

Thus, we have identified a simple mechanism (driven by the source term) of a rather striking 
and nontrivial phase transition in a generic fragmentation problem\cite{DM}. There is a physical 
meaning associated with this phase transition. For $m<m_c$, the fluctuation (variance) in the number of
splitting events scales as $N$ for large $N$ and the central limit theorem holds. In fact, one finds that
the full distribution of $n_N$ is Gaussian for $m<m_c$. However, for $m>m_c$, rare events 
give rise occasionally to huge fluctuations. In the language of the fragmentation problem, note that the effective 
distribution of the fraction $\eta(r)=(m-1)(1-r)^{m-2}$ gets highly localized around $r=0$ for large $m$.
This means that for large $m$, most of the $m$ fragments have very tiny lengths (which thus become `dead') except one 
which has
a huge length (due to the length conservation condition, $\sum_{i=1}^m r_i=1$). Thus this large piece
will persist for a long time and one will get a huge number of splitting events. This qualitative argument,
of course, does not explain why there is a sharp phase transition. For that, one has to carry out explicit calculations as 
done here. 

\section{Generalization to Vector Data String}

\begin{figure}[htbp]
\epsfxsize=8cm
\centerline{\epsfbox{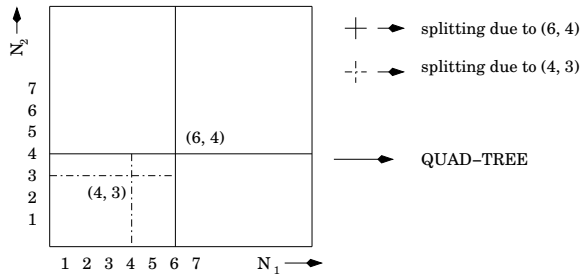}}
\caption{The storing of $(6,4)$ and $(4,3)$ on a quad tree $\rightarrow$ square fragmentation process.} 
\label{fig:sq1}
\end{figure}

So far, we have considered the storing of a data string of size $N$ on a tree where each element of the data is a scalar.
A natural generalization of this is when the data consists of a string of $N$ $D$-dimensional vectors.
For example, suppose we have the following data of $2$-dimensional vectors: $\{ (6,4), (4,3), (5,2), (8,7),\dots\}$.
How do we store this data on a tree? The corresponding tree is known as a quad-tree in the computer science 
literature\cite{FB}. To store this data, one imagines a $N\times N$ square. The first key $(6,4)$ is stored
at the co-ordinate $(6,4)$ of this square and it forms the root of the tree. This root has now $4$ branches
corresponding the $4$ quadrants around the point $(6,4)$. Note immediately the analogy to a corresponding
fragmentation problem. the storing of the first vector corresponds to fragmenting the original $N\times N$ square
into $4$ rectangles which join each other a $(6,4)$ (see Fig. \ref{fig:sq1}). This is the generalization of breaking a
one dimensional stick in the scalar case. Since both the components $6$ and $4$ are chosen independently and randomly
from the set $\{1,2,3,\dots,N\}$, this becomes a random fragmentation problem where the side lengths of any one of the 4
rectangles are chosen uniformly from the interval $[0-N]$.   

The next 
element $(4,3)$ arrives and storing $(4,3)$ is equivalent to the fragmentation of the
rectangle containing the new point $(4,3)$ into $4$ further smaller rectangles. This process
continues till all the data is stored, i.e., when the areas of all the rectangles become smaller
than some threshold value $A_0=1$. One immediately sees the generalization to the case where each
data element is a $D$-dimensional tuple. In the corresponding fragmentation problem, 
one starts with a $D$-dimensional cuboid of side lengths $N$ and the arrival of each data
corresponds to fragmenting a cuboid into $2^D$ number of smaller cuboids. Note that $D=1$
corresponds to the binary search tree of the scalar data, discussed before.

Following similar routes as in the $m$-ary search tree case, we were able to determine the exact asymptotic properties
of the height $H_N$, the balanced height $h_N$ and the number of non-empty nodes $n_N$ of a $D$-dimensional
quad-tree. We just mention our main results here without providing details since they are similar
to the earlier cases. For the extreme variables such as the height $H_N$ and the balanced height $h_N$, we again find a 
traveling front structure whose analysis provides us with the following exact asymptotics for large $N$,
\begin{eqnarray}
\langle H_N\rangle &=& 4.31107\dots \log N -\frac{1.95303\dots}{D} \log(D\log N) +\dots \nonumber \\
\langle h_N\rangle &\approx&  0.373365\dots \log N + \frac{0.89374\dots}{D} \log(D\log N) +\dots.
\label{quad11}
\end{eqnarray}
Surprisingly, the leading behavior (especially the coefficients of $\log (N)$ terms) turns out to be
independent of the dimension $D$. Besides, due to the existence of a traveling front structure, one immediately finds 
that the all
the higher moments including the variance of $H_N$ and $h_N$ are bounded $\sim O(1)$ for large $N$.
\begin{figure}[htbp]
\epsfxsize=8cm
\centerline{\epsfbox{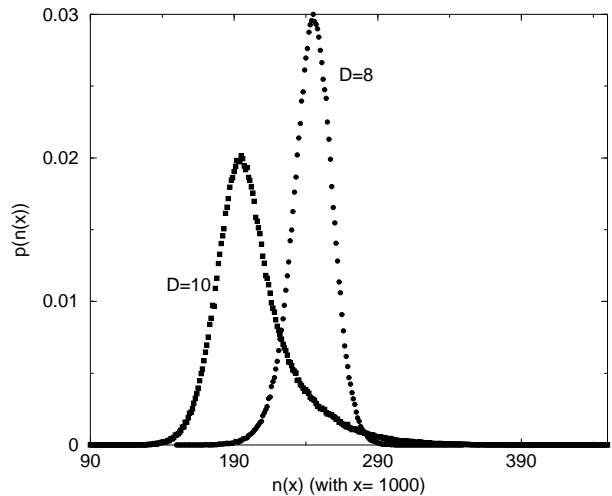}}
\caption{The distribution of the number of splittings of a cuboid with sidelength $N=x=1000$ for $D=8$ (filled circles) 
and for $D=10$ (filled squares). The distribution is Gaussian for $D=8$, but has a non-Gaussian skewness for $D=10$.
Note that the theoretically predicted critical dimension is $D_c=8.69363\dots$. The histogram was formed by numerically
splitting $5\times 10^5$ samples in each case.} 
\label{fig:hist}
\end{figure}

For the number of nodes $n_N$, we again find a phase transition\cite{DM} driven by the same mechanism mentioned 
earlier.
We find that while the average number of non-empty nodes $\mu(N)=\langle n_N\rangle \approx 2V/D$ for large $N$ where 
$V=N^D$, the variance $\nu(N)= \langle \left(n_N-\mu(N)\right)^2\rangle$ undergoes a phase transition at a critical value
of $D_c= \pi/\sin^{-1}\left(1/\sqrt{8}\right)=8.69363\dots$,
\begin{eqnarray}
\nu(N) &\sim & V \quad\quad {\rm for}\quad  D\le D_c \nonumber \\
       &\sim & V^{2\theta(D)} \quad\,\, {\rm for} \quad  D> D_c,
\label{quad22}
\end{eqnarray}
where $V=N^D$ and we computed the critical exponent $\theta(D)\ge 1/2$ exactly\cite{DM}
\begin{equation}
\theta(D)= 2\cos \left(\frac{2\pi}{D}\right)-1
\label{quad3}
\end{equation}
which increases monotonically with $D$ for $D>D_c$. Furthermore, we computed numerically the full distribution of $n_N$
for different values of $D$ and found that while the distribution is Gaussian for $D<D_c$ (a fact that can also be proved 
analytically), it becomes non-Gaussian for $D>D_c$ (see Fig. \ref{fig:hist}). As before, once we write down 
the moment equations, $D$ can be treated as a continuous parameter though in actual vector data $D$ represents 
the dimension of a vector element and therefore $D$ is always an integer.

\section{Conclusion}
In this paper, we have demonstrated how a variety of techniques developed in statistical physics can be successfully 
used to understand the statistical properties of various search trees, in particular for the random $m$-ary search tree 
problem. Search trees are the basic objects in
data storage and retrieval. Hence we expect that our results will have important consequences in 
the `sorting and searching' area of computer science. Our approach, perhaps not rigorous in the strict mathematical sense, 
has the advantage that it provides a physically transparent derivation of asymptotic results and can be readily 
generalized to study different types of search trees. For example, the traveling front method has subsequently been used 
to study the so called `digital search tree' that are used in the Lempel-Ziv data compression algorithm\cite{SM}.
Besides, our approach has the beauty that it makes links between seemingly different problems and provides
us with new results such as those for the vector data. We hope that the techniques discussed in this paper 
would be useful in future for studying other problems in
computer science.

\end{document}